\documentclass[12pt]{article}
\usepackage{amsfonts}
\usepackage{amsmath}
\usepackage{amssymb}
\usepackage{graphicx}
\usepackage{color}
\usepackage[all, knot]{xy}
\usepackage{tikz}

\usepackage[utf8]{inputenc}
\usepackage{epstopdf}
\usepackage[footnotesize]{caption}
\usepackage{amsthm}
\usepackage{enumitem}
\usepackage{mathrsfs}

\usepackage[margin=3cm]{geometry}

\def \be {\begin{equation}}
\def \ee {\end{equation}}
\def \bea {\begin{eqnarray}}
\def \eea {\end{eqnarray}}
\def \nn {\nonumber}

\def \rr {\raise.35ex\hbox{\small $\prime$}\kern-.17em{\mbox{\large $\imath$}}}

\def \dels {\partial\kern-.6em /\kern.1em}
\def \As {{A\kern-.5em / \kern.5em}}
\def \Ds {D\kern-.7em / \kern.5em}

\def \ks {k\kern-.5em /}
\def \ls {l\kern-.5em /}

\newcommand{\ci}[1]{}

\newcommand{\ba}{\begin{eqnarray}}
\newcommand{\ea}{\end{eqnarray}}
\newcommand{\bal}{\begin{align}}
\newcommand{\eal}{\end{align}}
\newcommand{\bay}[1]{\left(\begin{array}{#1}}
\newcommand{\eay}{\end{array}\right)}

\newcommand{\hide}[1]{}

\newtheorem{theorem}{Theorem}

\newtheorem{lemma}{Lemma}

\newlist{axioms}{enumerate}{2}
\setlist[axioms,1]{label=\textbf{A\arabic{axiomsi}.}, ref=A\arabic{axiomsi}}
\setlist[axioms,2]{label=\textbf{A\arabic{axiomsi}\rlap{\myEnumCounter{axiomsii}}.},%
                   ref=A\arabic{axiomsi}\myEnumCounter{axiomsii},%
                   align=parleft,%
                   leftmargin=0em,%
                   itemsep=1.4ex,%
                   before={\stepcounter{axiomsi}}}
                   
  \usetikzlibrary{decorations.markings}

\begin{document}

\begin{titlepage}

\begin{center}

\hfill
\vskip .2in

\textbf{\LARGE
Maximally Entangled State
and Bell's Inequality in Qubits 
\vskip.3cm
}
\vskip .5in
{\large
Su-Kuan Chu$^a$ \footnote{e-mail address: skchu@terpmail.umd.edu}, 
Chen-Te Ma$^b$ \footnote{e-mail address: yefgst@gmail.com}, 
Rong-Xin Miao$^c$ \footnote{e-mail address: mrx11@mail.ustc.edu.cn}\\ 
and Chih-Hung Wu $^b$ \footnote{e-mail address: b02202007@ntu.edu.tw}
\\
\vskip 1mm
}
{\sl
$^a$
Joint Quantum Institute and Joint Center for Quantum Information and Computer Science,
NIST/University of Maryland, College Park, Maryland 20742, USA,
\\
$^b$
Department of Physics and Center for Theoretical Sciences, \\
National Taiwan University, Taipei 10617, 
\\
$^c$
National Center for Theoretical Sciences, National Tsing-Hua University, \\
Hsinchu 30013, Taiwan, R.O.C..
}\\
\vskip 1mm
\vspace{40pt}
\end{center}
\begin{abstract}
A maximally entangled state is a quantum state which has maximum von Neumann entropy for each bipartition. Through proposing a new method to classify quantum states by using concurrences of pure states of a region, one can apply Bell's inequality to study intensity of quantum entanglement of maximally entangled states. We use a class of seven-qubit quantum states to demonstrate the method, where we express all coefficients of the quantum states in terms of concurrences of pure states of a region. When a critical point of an upper bound of Bell's inequality occurs in our quantum states, one of the quantum state is a ground state of the toric code model on a disk manifold. Our result also implies that the maximally entangled states does not suggest local maximum quantum entanglement in our quantum states.
\end{abstract}
\end{titlepage}

\section{Introduction}
\label{1}
Quantum entanglement is a physical phenomenon for that a quantum state of each particle cannot be described independently. When a quantum state in a quantum mechanical system has maximum von Neumann entropy for each bipartition, the entanglement entropy of a region $A$, $S_A\equiv-\mbox {Tr}_A \rho_A \ln \rho_A$, is equivalent to the Rényi entropy, $S_{\alpha, A}\equiv\ln\mbox{Tr}_A\rho_A^{\alpha}/(1-\alpha)$, of each order $\alpha$, where the reduced density matrix of the region A satisfies $\rho_A\equiv\mbox{Tr}_B \rho$. ($\mbox{Tr}_A$  is a partial trace operation on the region $A$ and $\mbox{Tr}_B$ is the same operation on the region $B$.)
We call such state having only one effective entanglement quantity---entanglement entropy--- a {\it maximally entangled state}. A maximally entangled state is useful for applications of quantum computing \cite{Steane:1997kb} and quantum algorithm \cite{Galindo:2001ei} as well as topological entanglement
entropy in a toric code model \cite{Kitaev:1997wr}, which was already realized in \cite{Li2016} by using a geometric algebra procedure \cite{Peng2010} or a combination of two-body interactions and radio-frequency pulses \cite{Peng2010, Tseng1999}. Topological entanglement entropy is a universal term of entanglement entropy in a toric code model \cite{Hamma:2005zz}. The toric code model also has a suitable tensor product decomposition of a region to extract topological entanglement entropy \cite{Hung:2015fla}. When a region is not contractible, the topological entanglement entropy should be lowered, but the toric model in a disk manifold with holes does not \cite{Chang:2017ygi}.

When quantum entanglement occurs, it distinguishes a quantum system from a classical system. The quantum nature of entanglement can be manifested \cite{Clauser:1969ny} through violation of Bell's inequality \cite{Bell:1964kc}, which states that correlations should satisfy the Bell's inequality when local realism is realized. The local realism is that information transmission cannot be faster than the speed of light and that there should be a pre-existing value before any possible measurement.

The violation of the Bell's inequality is measured in a two-qubit system and is also theoretically studied in \cite{Cirelson:1980ry}. The maximum violation of Bell's inequality in a two-qubit system is shown to be a monotonic function with respect to the concurrence of a pure state of a region $A$ \cite{Bennett:1996gf}, $C_A(\psi)\equiv\sqrt{2(1-\mbox{Tr}_A\rho_A^2)}$, and the entanglement entropy of the region $A$, $S_A$, through the two maximal eigenvalues of the $3 \times 3$ $R$-matrix $R_{ij}\equiv\mbox{Tr}(\rho\sigma_i\otimes\sigma_j)$ \cite{Verstraete:2001, Horodecki:1995},
where $i=1, 2, 3$ and $j=1, 2, 3$, and $\sigma_x$, $\sigma_y$, $\sigma_z$ are the Pauli matrices defined in \cite{PM}.
For some $n$-qubit quantum states, an upper bound of Bell's inequality is also a monotonic function with respect to the generalized concurrence of the pure state of a region $A$ \cite{Chang:2017ygi, Chang:2017czx}, $C_A(m, \psi)\equiv\sqrt{2(1-2^{m-1}\mbox{Tr}_A\rho_A^2)}$,
and the entanglement entropy of the region $A$  through the generalized $R$-matrix \cite{Chang:2017ygi, Chang:2017czx}, $R_{i_1i_2\cdots i_n}\equiv\mbox{Tr}(\rho\sigma_{i_1}\otimes\sigma_{i_2}\otimes\cdots\otimes\sigma_{i_n})\equiv R_{Ii_n}$,
where $i_\alpha=x,y,z$ and $\alpha=1, 2, \cdots, n$ are the site indices. 
The generalized $R$-matrix can be rewritten as a $3^{n-1} \times 3$ matrix $R_{I i_n}$ with the first index being a multi-index $I=i_1i_2\cdots i_{n-1}$ and the second index being $i_n$. The $n$-qubit Bell's operator ${\cal B}_n$ \cite{Gisin:1998ze} is
 \bea
{\cal B}_n&\equiv& A_1\otimes A_2\cdots\otimes A_{n-2}\otimes A_{n-1}\otimes \bigg(A_n+A_n^{\prime}\bigg)
\nn\\
&&+A^{\prime}_1\otimes A_2^{\prime}\cdots\otimes A_{n-2}^{\prime}\otimes A_{n-1}^{\prime}\otimes \bigg(A_n-A_n^{\prime}\bigg),
\label{Eq:mBell}
\eea
where $A_i\equiv {\bf a}_i \cdot \boldsymbol{\sigma}$ and $A^{\prime}_i\equiv {\bf a}_i^{\prime} \cdot \boldsymbol{\sigma}$ are the operators in the $i$-th qubit
with ${\bf a}_n$ and ${\bf a}_n^{\prime}$ being unit vectors and $\boldsymbol{\sigma}\equiv(\sigma_x,\sigma_y,\sigma_z)$ being the vector of Pauli matrices. Therefore, the $n$-qubit Bell's operator can offer a way to qualitatively study entanglement entropy.

In this letter, we discuss maximally entangled states in an $n$-qubit systems by using an upper bound of maximum violation of Bell's inequality. Quantum entanglement indicates whether a {\it quantum state} of each particle could be factorized from the state of a total system. Consequently, quantum entanglement is a fact about the full density matrix rather than a reduced density matrix. One special property of quantum entanglement is that information of the whole system can be partially extracted by observing one single particle of the quantum state. The higher the \textit{intensity} of quantum entanglement is, the stronger the correlation the system has. The upper bound of Bell's inequality provides a way to quantify this concept. Thus, we use an upper bound of Bell's inequality to discuss the intensity of quantum entanglement in an $n$-qubit maximally entangled state. Different maximally entangled states could have different intensity of quantum entanglement. One might also expect that in the Hilbert space the intensity of quantum entanglement of a maximally entangled state is at least a local maximum.  We discover that this can be a saddle point.

\section{Maximally Entangled State and Bell's Inequality}
\label{2}
We use seven-qubit quantum states as an example to estimate intensity of quantum entanglement in the maximally entangled state through Bell's inequality \cite{Bell:1964kc, Gisin:1998ze}. A ground state of a toric code model \cite{Kitaev:1997wr, Li2016} on a disk manifold can be shown to have local maximum quantum entanglement in our quantum states through an upper bound of Bell's inequality \cite{Chang:2017ygi, Verstraete:2001, Chang:2017czx}. We also discuss relations between an upper bound of Bell's inequality and concurrences of pure states of a region \cite{Bennett:1996gf}.

We choose the seven-qubit quantum states:
\bea
|\psi\rangle_7&\equiv&\alpha_1|000000\rangle_B|0\rangle_A+\alpha_2|000111\rangle_B|1\rangle_A
+\alpha_3|111100\rangle_B|0\rangle_A+\alpha_4|111011\rangle_B|1\rangle_A,
\nn\\
1&=&\alpha_1^2+\alpha_2^2+\alpha_3^2+\alpha_4^2.
\eea
As indicated above, the system is separated into two regions $A$ and $B$, where the region $B$ contains six qubits and the region $A$ contains one qubit. An upper bound of the Bell's inequality can be obtained by expressing $\alpha_1^2$, $\alpha_2^2$, $\alpha_3^2$ and $\alpha_4^2$ in terms of concurrences of pure states of the region $A$ \cite{Bennett:1996gf} (see Supplementary Material \cite{SM}). Firstly, we can write
$\alpha_1=\sin(\theta_1)$, $\alpha_2=\cos(\theta_1)\sin(\theta_2)$,
$\alpha_3=\cos(\theta_1)\cos(\theta_2)\cos(\theta_3)$, $\alpha_4=\cos(\theta_1)\cos(\theta_2)\sin(\theta_3)$,
where $0\le\theta_1\le 2\pi$, $0\le\theta_2,\theta_3, \theta_4\le \pi$. When we crank up $\theta_2=\pi/2$ and turn down $\theta_3=0$. Now we obtain new coefficients corresponding to the four product states $\beta_1=\sin(\theta_1)$, $\beta_2=\cos(\theta_1)$, $\beta_3=\beta_4=0$. Consequently, we get
$\mbox{Tr}_A\rho_{A}^2\equiv\mbox{Tr}_A\rho_{1,A}^2=\beta_1^4+\beta_2^4$.
We have the freedom to pick 
$\beta_1^2=1/2-\sqrt{1- C_{1, A}^2(\psi)}/2$, where 
$C_{1,A}(\psi)\equiv\sqrt{2\big(1-\mbox{Tr}_A\rho_{1, A}^2\big)}$.

Thirdly, we still choose $\theta_3=0$ but restore the value of $\theta_2$. We have another new coefficients
$\gamma_1=\sin(\theta_1)$, $\gamma_2=\cos(\theta_1)\sin(\theta_2)$, 
$\gamma_3=\cos(\theta_1)\cos(\theta_2)$, $\gamma_4=0$ and
$\mbox{Tr}_A\rho_{A}^2\equiv\mbox{Tr}_A\rho_{2, A}^2=(\gamma_1^2+\gamma_3^2)^2+\gamma_2^4$.
Hence, one solution is
$\cos^2(\theta_2)=\bigg(\sqrt{1-C_{1, A}^2(\psi)}+\sqrt{1-C_{2, A}^2(\psi)}\bigg)\bigg/\bigg(1+\sqrt{1-C_{1, A}^2(\psi)}\bigg)$,
where $C_{2, A}(\psi)\equiv\sqrt{2\big(1-\mbox{Tr}_A\rho_{2, A}^2\big)}$.

Finally, we reinstate both $\theta_2$ and $\theta_3$ in an attempt to express $\theta_3$ in terms of concurrences of pure states of the region $A$. The purity of the region $A$, $\mbox{Tr}_A\rho_A^2$, is $\mbox{Tr}_A\rho_A^2=(\alpha_1^2+\alpha_3^2)^2+(\alpha_2^2+\alpha_4^2)^2$.
Therefore, one solution is
\bea
\cos^2(\theta_3)
&=&\frac{\frac{\sqrt{1-C_{1, A}^2(\psi)}\sqrt{1-C_{2, A}^2(\psi)}+1-C_{1, A}^2(\psi)}{\sqrt{1-C_{1, A}^2(\psi)}+\sqrt{1-C_{2, A}^2(\psi)}}+\sqrt{1-C_A^2(\psi)}}{\sqrt{1-C_{1, A}^2(\psi)}+\sqrt{1-C^2_{2, A}(\psi)}}.
\eea
In so doing, we find that the quantum states can be classified by $C_{1,A}^2(\psi)$, which is determined by $\beta^2_1$, $C_{2, A}^2(\psi)$, which is determined by $\gamma^2_1$ and $\gamma^2_2$, and $C_A^2(\psi)$, which is determined by $\alpha^2_1$, $\alpha^2_2$ and $\alpha^2_3$. Here we do not give the most generic solution, but the most generic solution should be exactly solvable. We will demonstrate the upper bound of the Bell's inequality for the seven-qubit quantum states and show that a critical point of the upper bound of the Bell's inequality also corresponds to a ground state of the seven-qubit toric code model on a disk manifold since it is included in our solution.

To compute the upper bound of the Bell's inequality, which is defined by two largest eigenvalues of $R^{\dagger}R$, we introduce the lemma \cite{Chang:2017ygi, Horodecki:1995, Chang:2017czx}:
\begin{lemma}
\label{me}
The maximum violation of the Bell's inequalities $\gamma\equiv\max_{{\cal B}_n} {\rm Tr}(\rho{\cal B}_n)\le2\sqrt{u_1^2+u_2^2}$, where $u^2_1$ and $u^2_2$ are the first two largest eigenvalues of $R^\dagger R$ when $n>2$ and $\gamma=2\sqrt{u_1^2+u_2^2}$ when $n=2$.
\end{lemma}
By invoking the seven-qubit quantum states $|\psi\rangle_7$, we can show that $R^{\dagger} R$ only has diagonal elements (see Supplementary Material \cite{SM}):
$\big(R^{\dagger} R\big)_{xx}=\big(R^{\dagger} R\big)_{yy}=16(\alpha_1^2+\alpha_3^2)(\alpha_2^2+\alpha_4^2)+48(\alpha_1^2\alpha_4^2+\alpha_2^2\alpha_3^2)$
and
$\big(R^{\dagger}R\big)_{zz}
=1+32\alpha_1^2\alpha_3^2+32\alpha_2^2\alpha_4^2$,
and that a critical point of the upper bound of the Bell's inequality occurs at 
$\alpha_1^2=\alpha_2^2=\alpha_3^2=\alpha_4^2=1/4$ (see Supplementary Material \cite{SM}). For our solution, the critical point of the Bell's inequality corresponds to
$C_{1, A}^2(\psi)=C_{2, A}^2(\psi)=3/4$ and $C_A^2(\psi)=1$, which
also corresponds to maximally entangled states. The upper bound of the maximum violation of the Bell's inequality turns out to be $4\sqrt{5}$. The critical point also shows that the quantum states are not locally maximally entangled because the critical point of the upper bound of the Bell's inequality is a saddle point. 

If we consider $C_{1,A}^2(\psi)=C_{2, A}^2(\psi)=C_A^2(\psi)=1$,  the coefficients are $\alpha_1^2=\alpha_2^2=1/2$ and $\alpha_3^2=\alpha_4^2=0$. The upper bound of the Bell's inequality is $4\sqrt{2}< 4\sqrt{5}$, which can be shown from the following theorem \cite{Chang:2017ygi, Verstraete:2001, Chang:2017czx}:
\begin{theorem}
\label{mec}
For an $n$-qubit quantum state $|\psi \rangle=|u\rangle_B\otimes\big(\lambda_+|v\rangle_B \otimes |1\rangle_A +\lambda_-|\tilde{v}\rangle_B \otimes |0\rangle_A\big)$
with $\lambda_+|v\rangle_B \otimes |1\rangle_A +\lambda_-|\tilde{v}\rangle_B \otimes |0\rangle_A$ being a non-biseparable $2\alpha$-qubit state, where $\alpha$ is an integer,
$|u\rangle_B$, $|v\rangle_B$, $|\tilde{v}\rangle_B$ being product states consisting of $| 0 \rangle$'s and $| 1 \rangle$'s, $|v\rangle_B$ and $|\tilde{v}\rangle_B$ are orthogonal, and
the maximum violation of the Bell's inequality $\gamma$ in an $n$-qubit system is $\gamma\equiv\max_{{\cal B}_n} {\rm Tr}(\rho{{\cal B}}_n)=2f_{2\alpha}(\psi)$,
in which the function $f_{2\alpha}(\psi)$ is defined as:
$f_{2\alpha}(\psi)\equiv\sqrt{1+2^{2\alpha-2}C_A^2(\psi)}$ when $2^{2-2\alpha}\ge C_A^2(\psi)$,
$f_{2\alpha}(\psi)\equiv2^{\frac{2\alpha-1}{2}}C_A(\psi)$ when $2^{2-2\alpha}\le C_A^2(\psi)$.
The coefficients $\lambda_{\pm}$ satisfy $\lambda_{+}^2=\big(1\pm\sqrt{1-C_A^2(\psi)}\big)/2$ and $\lambda_{-}^2=\big(1\mp\sqrt{1-C_A^2(\psi)}\big)/2$.
\end{theorem} 
Because the first three sites of the quantum states are not entangled and other sites of the quantum states are maximally entangled, the upper bound of the Bell's inequality is equivalent to the maximum upper bound of the maximum violation of the Bell's inequality of a four-qubit system. It is worthy to note that even though the concurrences of the pure states $C_{1,A}^2(\psi)$, $C_{2, A}^2(\psi)$, and $C_A^2(\psi)$ are maximal, the upper bound of the Bell's inequality $4\sqrt{2}$ is smaller than that in the previous case of the upper bound of the Bell's inequality $4\sqrt{5}$, in which all coefficients have equal amplitude.

\begin{figure}
\centering{}\includegraphics[scale=0.32]{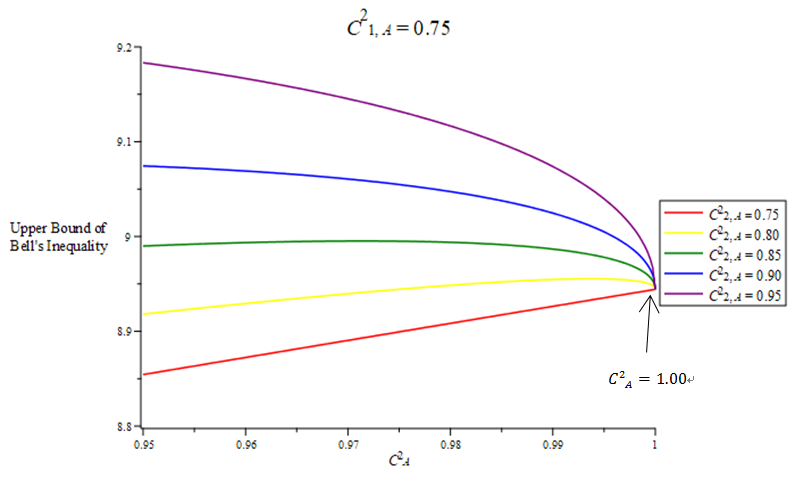}\caption{We fix $C_{1,A}^2=0.75$ and choose $C_{2,A}^2=0.75,\ 0.8,\ 0.85,\ 0.9,\ 0.95$ to compute the upper bound of maximum violation of the Bell's inequality. It is interesting to note that the upper bound of maximum violation of the Bell's inequality is independent of $C_{2,A}^2$ when $C_A^2=1$.\label{C1} }
\end{figure}

\begin{figure}
\centering{}\includegraphics[scale=0.32]{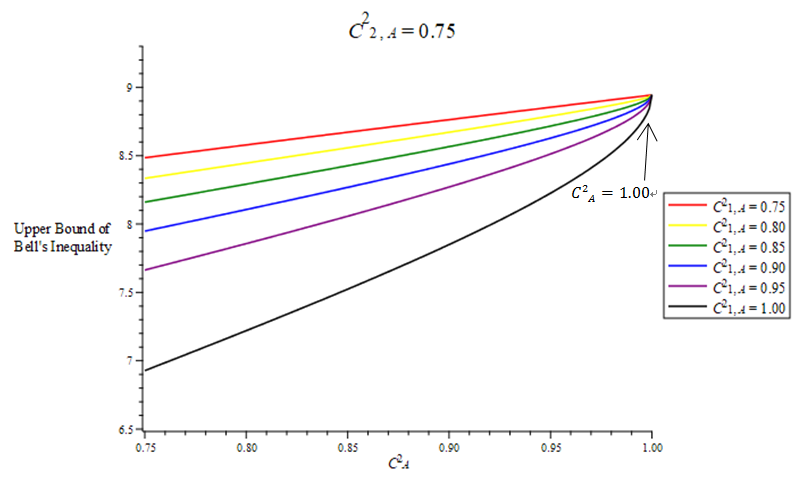}\caption{We fix $C_{2,A}^2=0.75$ and choose $C_{1,A}^2=0.75,\ 0.8,\ 0.85,\ 0.9,\ 0.95,\ 1$ to compute the upper bound of maximum violation of the Bell's inequality. It is interesting to note that the upper bound of maximum violation of the Bell's inequality is independent of $C_{1,A}^2$ when $C_A^2=1$. \label{C3} }
\end{figure}

Note that the upper bound of the Bell's inequality is independent of $C_{2,A}^2$ when the seven-qubit quantum states satisfy  $1-2\sqrt{1-C_{1,A}^2(\psi)}-\sqrt{1-C_A^2(\psi)}=0$ and $R_{xx}\ge R_{zz}$. The property is demonstrated in Fig. \ref{C1}. 
When the seven-qubit quantum states satisfy $1-2\sqrt{1-C_{2, A}^2(\psi)}+\sqrt{1-C_A^2(\psi)}=0$ and $R_{xx}\ge R_{zz}$, the upper bound of the Bell's inequality is also independent of $C_{1,A}^2$. The property is also demonstrated in Fig. \ref{C3}. 
When we prepare the maximally entangled state $\alpha_1^2=\alpha_2^2=\alpha_3^2=\alpha_4^2=1/4$ in an experiment \cite{Li2016}, this property can give an additional consistency check.

Although entanglement entropy of the region $A$ is also a monotonic function with respect to $C_A^2(\psi)$ as in the case of two qubits \cite{Verstraete:2001}, the upper bound of the Bell's inequality in general is not. This is shown in Fig. \ref{C1} 
and Fig. \ref{C3}. 
The non-monotonic property reflects that a system with a higher number of qubits can have more interesting entanglement structure than a two-qubit quantum system.

\begin{figure}
\centering{}\includegraphics[scale=0.2]{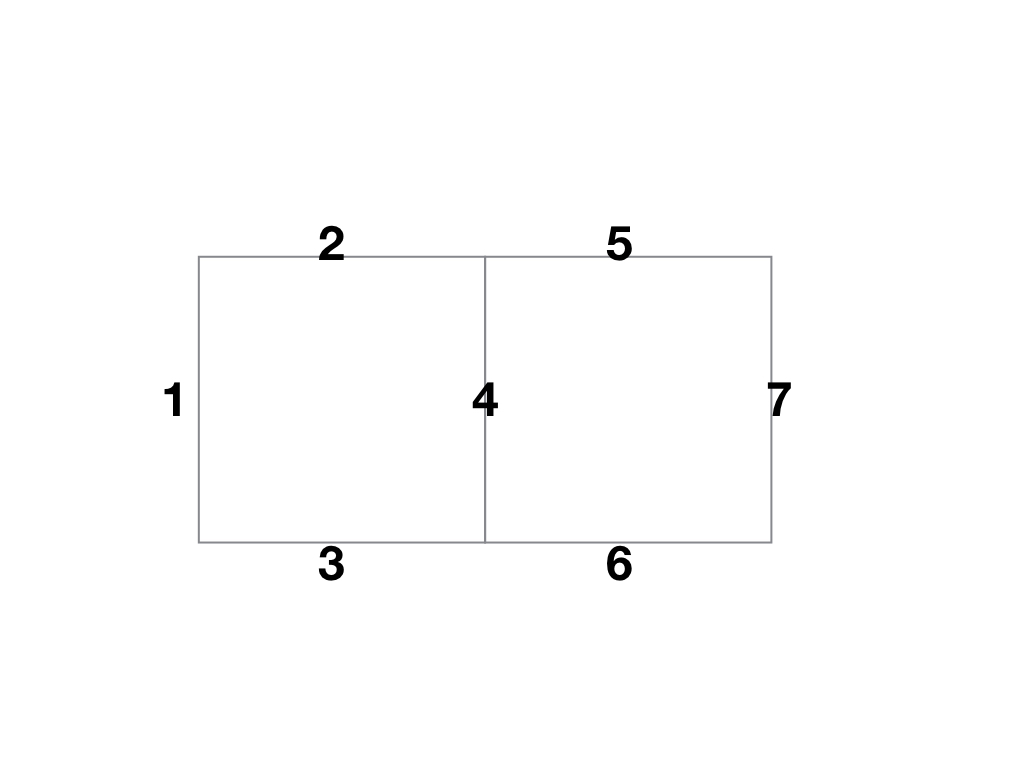}\caption{The sites of the qubits of the seven-qubit toric code model on a disk manifold are labeled by the integers. The boundaries are not identified on the disk manifold. \label{fig:7qubitdisk} }
\end{figure}

When the seven-qubit quantum state $|\psi\rangle_7$ demonstrates a critical point of the upper bound of the Bell's inequality, one quantum state is a ground state of a seven-qubit toric code model on a disk manifold. The Hamiltonian of the seven-qubit toric code model on a disk manifold is defined as
$H_{7td}\equiv-\sum_{i=1}^6A_i-\sum_{j=1}^2B_j$,
where
$A_{1}=\sigma_{z}\otimes\sigma_{z}\otimes 1\otimes 1\otimes 1\otimes 1\otimes 1$,  
$A_{2}=\sigma_{z}\otimes 1\otimes\sigma_{z}\otimes 1\otimes 1\otimes 1\otimes 1$,
$A_{3}=1\otimes\sigma_{z}\otimes 1\otimes\sigma_{z}\otimes\sigma_{z}\otimes 1\otimes 1$, 
$A_{4}=1\otimes 1\otimes\sigma_{z}\otimes\sigma_{z}\otimes 1\otimes\sigma_{z}\otimes 1$,
$A_{5}=1\otimes 1\otimes 1\otimes 1\otimes\sigma_{z}\otimes 1\otimes\sigma_{z}$, 
$A_{6}=1\otimes 1\otimes 1\otimes 1\otimes 1\otimes\sigma_{z}\otimes\sigma_{z}$,
$B_{1}=\sigma_{x}\otimes\sigma_{x}\otimes\sigma_{x}\otimes\sigma_{x}\otimes 1\otimes 1\otimes 1$, 
$B_{2}=1\otimes 1\otimes 1\otimes\sigma_{x}\otimes\sigma_{x}\otimes\sigma_{x}\otimes\sigma_{x}$.
The sites of the lattice are labeled as in Fig. \ref{fig:7qubitdisk}.

\section{Outlook}
\label{3}
Our goal is to show the intensity of quantum entanglement in a maximally entangled state by using violation of Bell's inequality \cite{Chang:2017ygi, Chang:2017czx, Gisin:1998ze}. This could help us to know the physical meaning for the definition of the maximum entangled state. Our theoretical result can also be realized in experiments \cite{Li2016}.

\section*{Acknowledgments}
Chen-Te Ma would like to thank Nan-Peng Ma for his encouragement. Su-Kuan Chu was supported by AFOSR, NSF QIS, ARL CDQI, ARO MURI, ARO and NSF PFC at JQI and Rong-Xin Miao was supported in part by NCTS and the grant MOST 105-2811-M-007-021 of the Ministry of Science and Technology of Taiwan.

\end{document}